\newcommand{\CLASSINPUTtoptextmargin}{2.4cm}
\newcommand{\CLASSINPUTbottomtextmargin}{4.6cm}
\documentclass[conference]{IEEEtran}
\IEEEoverridecommandlockouts
\usepackage{cite}
\usepackage[left=1.62cm,right=1.62cm,top=1.9cm]{geometry}

\usepackage{amsmath,amssymb,amsfonts}
\usepackage{mdwmath}
\usepackage{algorithm, algorithmic}
\usepackage{graphicx}
\usepackage{mathabx}
\usepackage{mdwtab}
\usepackage{textcomp}
\usepackage{xcolor}
\usepackage{booktabs}
\usepackage{graphicx}
\usepackage{enumitem}
\usepackage{harpoon,amsmath}

\usepackage[left=1.62cm,right=1.62cm,top=1.9cm]{geometry}

\usepackage[explicit]{titlesec}
\usepackage{subfigure}
\usepackage[numbers,sort&compress]{natbib} 
\def\BibTeX{{\rm B\kern-.05em{\sc i\kern-.025em b}\kern-.08em
    T\kern-.1667em\lower.7ex\hbox{E}\kern-.125emX}}
\IEEEoverridecommandlockouts
\begin{document}
\titlespacing{\subsection}{0pt}{0.1ex plus 1ex minus .1ex}{0.2ex plus .2ex}
\titlespacing{\section}{0pt}{0.1ex plus 1ex minus .1ex}{0.2ex plus .2ex}

\title{
Goal-oriented Semantic Communication for the Metaverse Application\\
}

\author{\IEEEauthorblockN{Zhe Wang\IEEEauthorrefmark{1}, Nan Li\IEEEauthorrefmark{1}, Yansha Deng\IEEEauthorrefmark{1}}
\IEEEauthorblockA{King's College London\IEEEauthorrefmark{1} 
\\
\{tylor.wang, nan.3.li, yansha.deng\}@kcl.ac.uk}}

\maketitle

\begin{abstract}
With the emergence of the metaverse and its role in enabling real-time simulation and analysis of real-world counterparts, an increasing number of personalized metaverse scenarios are being created to influence entertainment experiences and social behaviors. 
However, compared to traditional image and video entertainment applications, the exact transmission of the vast amount of metaverse-associated information significantly challenges the capacity of existing bit-oriented communication networks.
Moreover, the current metaverse also witnesses a growing goal shift for transmitting the meaning behind custom-designed content, such as user-designed buildings and avatars, rather than exact copies of physical objects.
To meet this growing goal shift and bandwidth challenge, this paper proposes a goal-oriented semantic communication framework for metaverse application (GSCM) to explore and define semantic information through the goal levels.
Specifically, we first analyze the traditional image communication framework in metaverse construction and then detail our proposed semantic information along with the end-to-end wireless communication.
We then describe the designed modules of the GSCM framework, including goal-oriented semantic information extraction, base knowledge definition, and neural radiance field (NeRF) based metaverse construction.
Finally, numerous experiments have been conducted to demonstrate that, compared to image communication, our proposed GSCM framework decreases transmission latency by up to 92.6\% and enhances the virtual object operation accuracy and metaverse construction clearance by up to 45.6\% and 44.7\%, respectively.
\end{abstract}


\begin{figure*}[t]
  \centering
  \includegraphics[width=0.87\textwidth ]{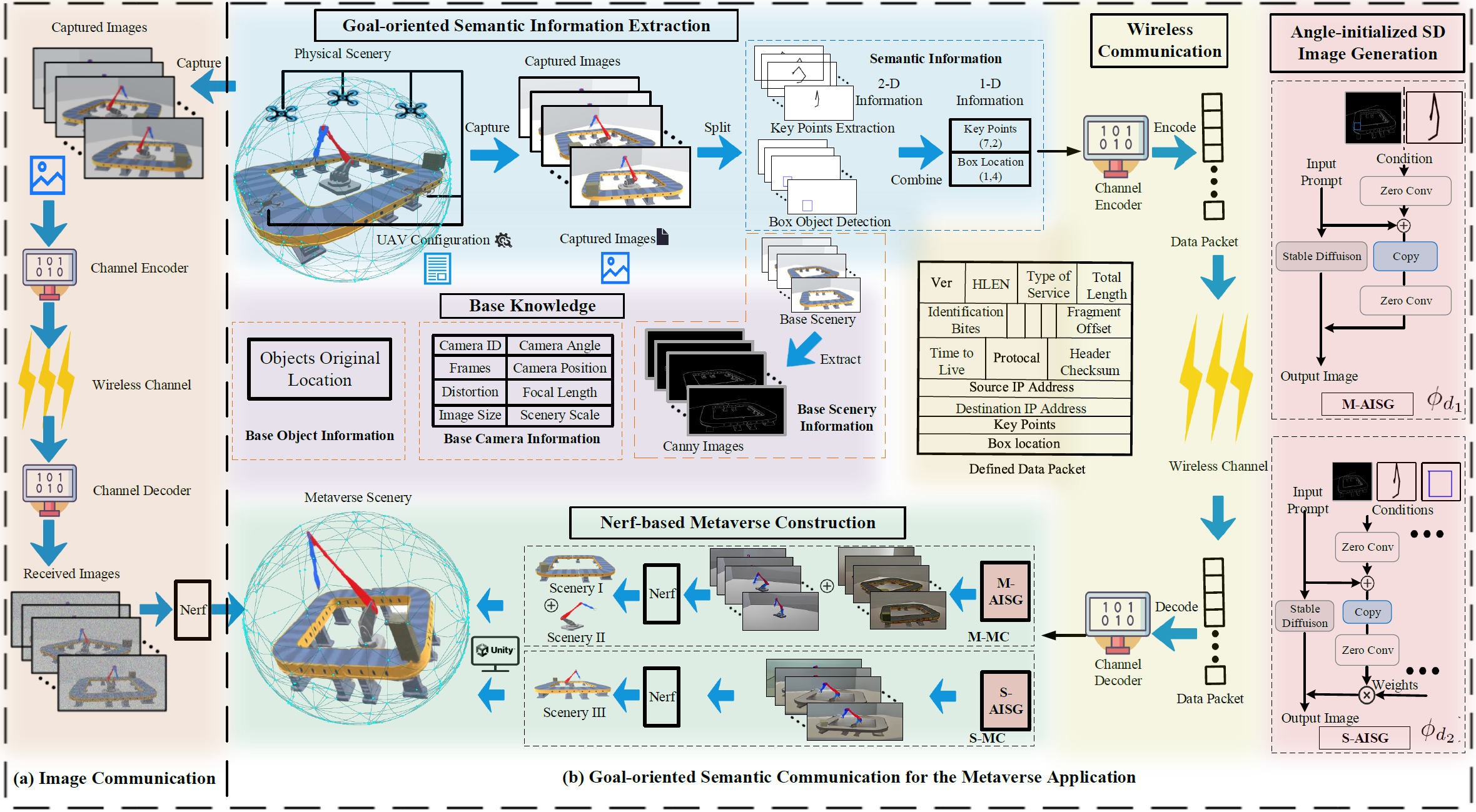} 
  \caption{Image communication framework and goal-oriented semantic communication framework for metaverse application} 
  \label{wirelesscommunication} 
\end{figure*}

\section{Introduction}
Metaverse, as an extension of the digital universe, incorporates applications such as gaming, social activities, real-world analysis, and simulation. 
Compared to research in one and two dimension information, such as text and image, metaverse applications incorporate three-dimensional data and need to process rich and complex data involving wireless communication between the transmitter and the edge, thus requiring stricter bandwidth restrictions \cite{wang2022survey}.
Additionally, metaverse applications require a comprehensive understanding of physical objects to simulate virtual counterparts accurately and create immersive environments for clients \cite{cao2023toward}. 
This, in turn, necessitates higher data transmission rates to handle the detailed physical object information needed for metaverse construction.
Previous research has demonstrated that metaverse clients typically require a bandwidth of about 5.6 Gbps for raw metaverse data communication \cite{dong2022metaverse}. However, the average 5G global wireless download bandwidth is only about 160 Mbps, posing significant challenges for the development of metaverse applications, especially in metaverse applications that involve real-time communication and interaction \cite{kumar20235g}.

Current research in metaverse construction focuses on the precise rendering of visual content from physical counterparts through wireless communication. 
To achieve this, the wireless metaverse applications concentrate on accurately transmitting various forms of data, such as point clouds, meshes, and $360^\circ$ videos, through bit-oriented communication \cite{pan2023litrix}.
For instance, in \cite{hu2021cooperative}, real-time $360^\circ$ video has been captured and transmitted from UAVs to VR users in a large sports event.
However, current metaverse applications are witnessing a goal shift towards incorporating more user-customized content \cite{zhou2024goaloriented}. The social aspects of the metaverse, such as user-designed scenery, buildings, and avatars, emphasize the transmission of meaning rather than exact replicas of physical objects \cite{wang2023task}.
Previous research has demonstrated that using custom scenery and avatar could encourage users to complete tasks more quickly in gaming scenarios \cite{ham2024virtual}. 
Thus, exact avatar skin colors or architectural decorations may be less critical in the metaverse construction, where diverse clients' beauty tastes within their custom digital metaverse.
Therefore, it is essential to redesign the communication framework to overcome bandwidth limitations while accounting for the specific goals of metaverse applications, moving beyond traditional metrics like accurate construction that focus on exact data replication to custom content representation.
The metaverse features extensive virtual objects, with each requiring significant amounts of data to be rendered, to create a vast and immersive virtual environment.
Traditionally, constructing 3D scenery requires numerous input images to render the scenery accurately. Generally, each frame necessitates the transmission of 1.5 Gb of graphics data for multiple virtual objects scenery construction \cite{mehrabi2021multi}.
Recent 3D rendering technologies have witnessed significant progress. 
For instance, the Neural Radiance Field (NeRF) has been proposed to map a 3D scenery using relatively fewer image inputs, thus reducing computational complexity and rendering time \cite{gao2022nerf}.
However, NeRF still encounters problems such as relying solely on images to accurately construct virtual counterparts and requiring a large number of image inputs for constructing large landscape 3D scenery.
This limitation hinders the development of the metaverse application, as precise replication of physical objects still cannot satisfy the requirements for client custom digital content.
Existing artificial intelligence generation information (AGI) technologies, such as stable diffusion (SD) and large language models (LLM), have demonstrated significant capabilities in image and text generation. These AGI technologies could potentially expand clients' inspiration and assist with custom metaverse construction \cite{sun2024spatial}. 
However, few researchers have incorporated AGI technologies within the metaverse, and maintaining SD image generation within the same scenery still remains an unsolved challenge.

To address these above limitations, we propose a goal-oriented semantic communication for the metaverse applications. In contrast to image communication framework that rely solely on image transmission, our proposed GSCM extracts both 1-D and 2-D semantic information, requiring less bandwidth while maintaining communication accuracy. The contributions of this paper can be summarized as:

\begin{enumerate}[itemsep=0pt]
\item We propose a goal-oriented semantic communication framework for the metaverse application to construct and simulate real-world scenery. The framework includes goal-oriented semantic information extraction, base knowledge definition, and NeRF-based metaverse construction.

\item We employ an innovative angle-initialized SD image generation (AISG) algorithm integrated with the NeRF rendering algorithm to simultaneously generate custom metaverse scenery.
Additionally, by utilizing base knowledge, we demonstrate the effectiveness of AISG in improving virtual object operation while maintaining metaverse construction clearance after wireless communication.

\item 
We have conducted a series of experiments to compare our proposed GSCM with the traditional image communication framework. Our results indicate that the GSCM framework significantly outperforms the image communication framework in object operation, scenery clearance, and transmission latency for wireless metaverse construction, showing improvements of 45.6\%, 44.7\%, and 92.6\%, respectively.

\end{enumerate} 

The rest of the paper is organized as follows: Section II presents the system model and problem formation.
Section III details the proposed GSCM framework.
Section IV outlines the evaluation metrics and experiment performance.
Finally, Section V concludes the paper.


\section{\small{System Model and Problem Formation}}
In this section, we first describe the existing image communication framework (ImageCom) shown in Fig. \ref{wirelesscommunication} (a). Then, we introduce our proposed GSCM, as depicted in Fig. \ref{wirelesscommunication} (b), which considers both the bit-level and the semantic and goal levels. We further present our wireless communication channel implemented in both the ImageCom and the GSCM frameworks. Finally, we present the problem formation.

\subsection{Image Communication Framework}
We utilize a benchmark image communication framework detailed in \cite{wang2024mulan}, which includes image capturing, wireless communication, and NeRF scenery construction.
With a set of evenly distributed, fixed-location UAVs $\mathcal{U}$ around a factory scenery. 
At the time slot $t$,  each UAV captures its image from its orientation. The image set $\mathcal{V}_t$ can be represented as
\begin{equation}
\mathcal{V}_t=[\mathbf{I}_1 \cdots, \mathbf{I}_{\text{N}_{\text{u}}}]^{\text{T}}, \quad \forall \mathbf{I}_i \in \mathbb{R}^{1200 \times 600 \times 3},\quad  i \in [1, \text{N}_{\text{u}}],
\end{equation}
where $\text{N}_{\text{u}}$ denotes the total number of UAVs, and $\mathbf{I}_i$ represents the RGB image matrix captured by the $i$-th UAV, with each image having a pixel resolution of $(1200\times600\times3)$. 
At the edge side, the received image set $\mathcal{V}^{\prime}_t$ after wireless communication, along with the UAV angle parameters, are utilized to construct the 3D scenery using a general NeRF algorithm. The NeRF algorithm output is calculated in point cloud formate $\mathbf{P}_{\text{c}}$ as
\begin{equation}
\label{construct}
\mathbf{P}_{\text{c}}=[\overrightharp{v}_{1}, \overrightharp{v}_{2}, \cdots, \overrightharp{v}_{\text{N}_{\text{c}}} ]^\text{T}=\mathcal{R}\left(\mathcal{V}^{\prime}_t, \mathcal{C}, \delta_{\text{r}} \right),
\end{equation}
where UAV configuration set $\mathcal{C}$ includes camera distortion, field of view (FoV), etc., $\delta_{r}$ represents NeRF algorithm neural network parameters, $\text{N}_{\text{c}}$ represents the number of points in the 3D scenery, and each point vector $\overrightharp{v}_{i}$ includes a three-dimensional location and RGB color information.


\subsection{Goal-oriented Semantic Communication Framework }
The proposed GSCM, based on the ImageCom framework, incorporates semantic and goal-level optimization within the factory scenery. In this scenery, virtual objects can be categorized into movable objects $\mathcal{M}$ and stationary background $\mathcal{O}$.
\subsubsection{Goal-oriented Semantic Information Extraction}
The moving object $\mathcal{M}$ undergoes pose and position change with metaverse operation and thus forms a contributing part of metaverse operational status representation. 
In contrast, the stationary background $\mathcal{O}$, while also an important component of the metaverse, generally does not experience posture or position changes with time. 
With the timeliness of motion information updates reflects the precision of a metaverse representing physical scenery.
The goal-oriented semantic information extraction $\mathcal{S}{(\cdot)}$ is defined to extract the information $\mathbf{S}_t$ representing object pose and position at each time $t$, such as key points and location, from the image set, which is represented as
\begin{equation}
\label{semantic_ext}
\mathbf{S}_t = [ \overrightharp{K}_t,  \overrightharp{B}_t]^\text{T}  =[k_1 \cdots, k_{\text{N}_\text{k}}, b_1 \cdots,  b_{\text{N}_\text{b}}]^\text{T} =\mathcal{S}\left(\mathcal{V}_t, \theta_{\mathrm{s}}\right),
\end{equation}
where 
$\overrightharp{K}_t$ represents the key points information vector, $\overrightharp{B}_t$ represents the location and size information vector, and $\theta_{\mathrm{s}}$ represents the neural network parameters, $\text{N}_\text{k}$ and $\text{N}_\text{b}$ represent the total number of parameters in $\overrightharp{K}_t$ and $\overrightharp{B}_t$, respectively.

\subsubsection{Nerf-based Metaverse Construction}
Compared with traditional ImageCom framework that rely solely on the received image set $\mathcal{V}_t^{\prime}$ for scenery construction, our proposed NeRF-based metaverse construction depends on defined semantic information, base knowledge, and prompt text descriptions to build metaverse scenery. 
The base knowledge includes the unchanged information in the metaverse that can be transmitted at time $t$ and will be detailed in section III.B.
To begin with, the edge side decodes the semantic information $\mathbf{S}^{\prime}_t$ with the received binary code after the wireless communication. 
Then the AISG algorithm $\mathcal{D}(\cdot)$ is initialized to regenerate the $\text{N}_\text{u}$ images with object operation status from varies angles. The regenerated images set $\hat{\mathcal{V}}_t^{\prime}$ is represented as
\begin{equation}
\hat{\mathcal{V}}_t^{\prime}= [\hat{\mathbf{I}}_1^{\prime} \cdots, \hat{\mathbf{I}}_{\text{N}_\text{u}}^{\prime}]^\text{T}=\mathcal{D}\left( \mathbf{S}^{\prime}_t, \mathcal{B}, W, \phi_\text{d}\right),
\end{equation}
where $\mathcal{B}$ represents the defined base information, $W$ represents the prompt text description, and $\phi_\text{d}$ represents the neural network parameters of the AISG algorithm. 
Finally, with the regenerated images set $\hat{\mathcal{V}}_t^{\prime}$ and UAV configuration $\mathcal{C}$ as input, the NeRF algorithm is employed to construction the metaverse scenery in point cloud $\mathbf{P}_{\text{m}}$ by Eq. (\ref{construct}).

\subsection{Wireless Communication}
Both the captured images from the ImageCom framework and the semantic information $\mathbf{S}_t$ from the GSCM framework are transmitted to the edge via a same wireless communication channel. 
Binary code information vector $\overrightharp{c}_t$ encoded by either image $\mathbf{I}_i$ or $\mathbf{S}_t$ with power $P=\mathbb{E}({|\overrightharp{c}_t|^2})$ is transmitted through the communication channel. We model the communication channel between a single-antenna transmitter and edge in a wireless environment based on an additive white Gaussian noise (AWGN) and a direct line-of-sight Rayleigh fading channel gain. The received data $\overrightharp{r}_t$ is represented as
\begin{equation}
\overrightharp{r}_t= \overrightharp{h}_t \otimes \overrightharp{c}_t+\overrightharp{w}_t,
\end{equation}
where $\overrightharp{w} \sim \mathcal{N}_{\mathbb{C}}\left(0, {\sigma_n}^2\right)$ denotes the AWGN with variance $\sigma_n^2$, and $\overrightharp{h}$ represents the channel gain vector of the direct path characterized by Rayleigh fading.
For a single transmission at time $t$, the Signal-to-Noise Ratio (SNR) is calculated with the signal power $P$ and noise, which is detailed as 
\begin{equation}
\text{SNR} = \frac{|\overrightharp{h}_t|^2 P}{{\sigma_n}^2}= \frac{\mathbb{E}({|\overrightharp{c}_t|^2}) |\overrightharp{h}_t|^2}{{\sigma_n}^2},
\end{equation}
where ${\sigma_n}^2$ represents the AWGN noise power.

\subsection{Problem formation}
The goal of the proposed GSCM framework, as compared to previous exact replicas of the physical world in ImageCom framework, is to create a 3D scenery that precisely reflects the operation status of physical object, such as pose and position. Thus, the objective function of GSCM is designed to minimize the gesture and location errors between the observed and constructed objects, which is represented as
\begin{equation}
\begin{aligned}
\label{pf}
& \arg \min_{(\theta_{\text{s}}, \phi_{\text{d}},\delta_{\text{r}})} \sum_{t=0}^{T} \frac{1}{T}  \mathcal{L}(\mathcal{S}(\hat{{\mathcal{V}_t}}) , {\mathcal{S}(\mathcal{V}_t}))
 \\ &= \arg \min_{(\theta_{\text{s}}, \phi_{\text{d}},\delta_{\text{r}})} \sum_{t}^{T} \frac{1}{T} {\frac{\sum_{1}^{\text{N}_\text{k}} (k_i-\hat{k}_i)+\sum_{1}^{\text{N}_\text{b}} (b_i-\hat{b}_i)}{{\text{N}_\text{k}}+{\text{N}_\text{b}}}},
\end{aligned}
\end{equation}
where $\hat{k}_i$ and $\hat{b}_i$ represent key points, position, and size information extracted from the images at the edge side.

\section{Goal-oriented Semantic Communication}

\subsection{Goal-oriented Semantic Information Extraction}
To precisely extract the operational status of metaverse, the semantic information extraction module involves two steps: 
1) A key points extraction algorithm that extracts 7 two-dimensional vector coordinates $\overrightharp{K}_t$ representing the pose of a robotic arm. 
2) A box object detection algorithm that extracts a four-parameter vector for a moving box $\overrightharp{B}_t$ representing the location and size of a moving box.

\begin{figure}[t]
  \centering
  \includegraphics[width=0.48\textwidth ]{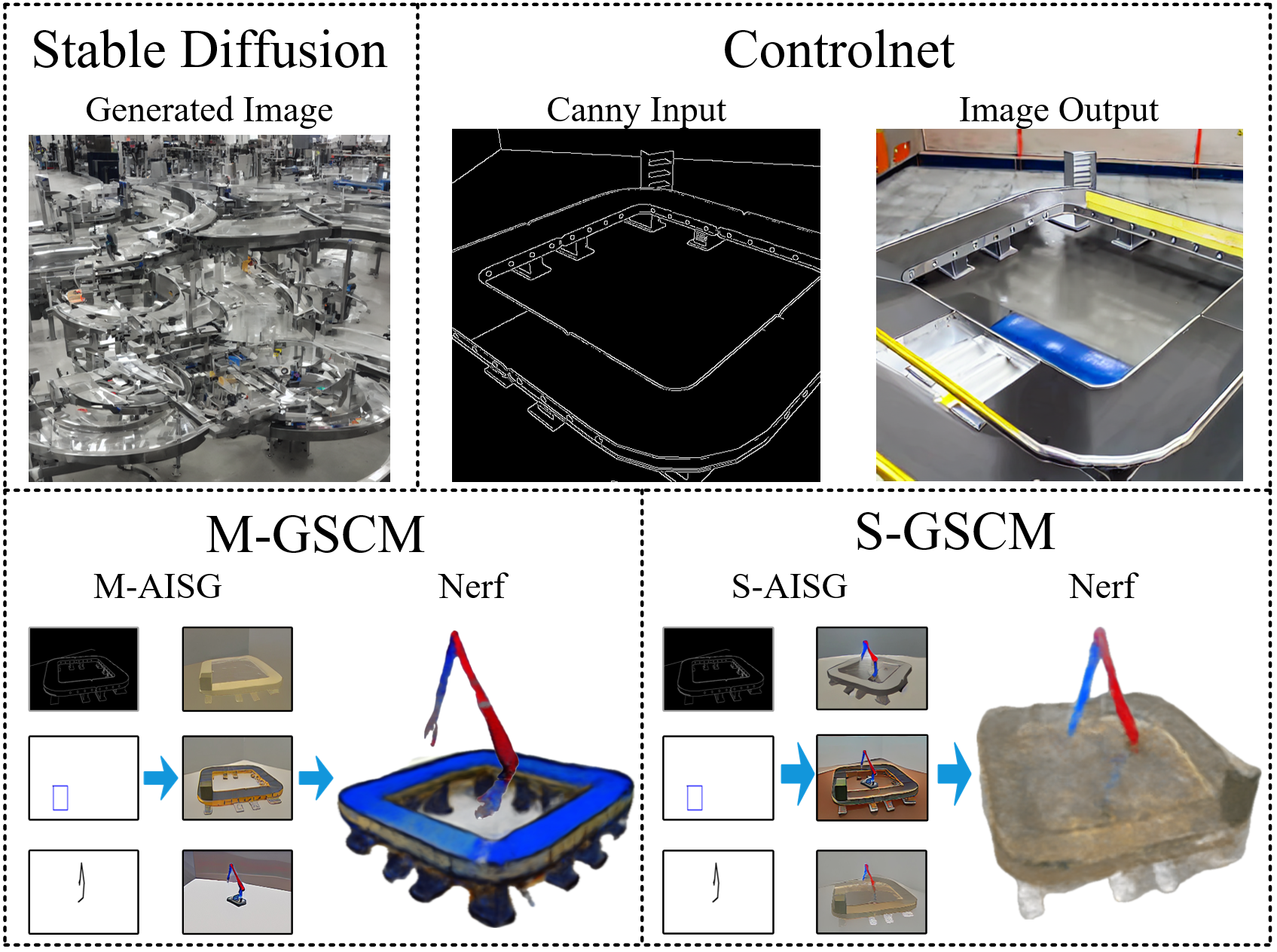} 
  \caption{Nerf-based metaverse construction} 
  \label{image_generation} 
\end{figure}

\textbf{Key Points Extraction Algorithm}: The robotic arm's key point is extracted by a calculation of UAV's pose configuration and the robotic arm coordinates in the Unity3D, which involves converting Unity3D coordinates to UAV relative coordinates and converting UAV relative coordinates into UAV image coordinates.
The key points converting from Unity3D coordinates $\overrightharp{P}_w$ to the UAV view coordinate $\overrightharp{P}_v$ is represented as
\begin{equation}
\overrightharp{P}_v = \mathbf{V} \cdot \overrightharp{P}_w, 
\end{equation}
where the homogeneity UAV view matrix $\mathbf{V}=\begin{bmatrix}
\mathbf{R} & \overrightharp{P}_u \\
0 & 1
\end{bmatrix}^{\text{T}}$ 
includes a $(3 \times 3)$ UAV rotation matrix $\mathbf{R}$ and a $(3 \times 1)$ vector of the UAV's position $\overrightharp{P}_u$ in Unity3D.
The key points converting from UAV relative coordinates $\overrightharp{P}_v$ to UAV image coordinates $\overrightharp{K}$ is represented as 
\begin{equation}
\vec{K}=\mathbf{M} \cdot \vec{P}_v,
\end{equation}
where $\mathbf{M}$ is the perspective projection matrix, incorporate the depth and spatial relationships from UAV camera lens to the 3D scenery, which is represented as
\begin{equation}
\mathbf{M}=\left[\begin{array}{cccc}
\frac{f_l}{a} & 0 & 0 & 0 \\
0 & f_l & 0 & 0 \\
0 & 0 & \frac{z_f+z_{\mathrm{n}}}{z_{\mathrm{n}}-z_f} & \frac{2 \times z_{\mathrm{f}} \times z_{\mathrm{n}}}{z_{\mathrm{n}}-z_{\mathrm{f}}} \\
0 & 0 & -1 & 0
\end{array}\right]^{\mathrm{T}}
\end{equation}
where $a$ is the aspect ratio between the width and height of the UAV image, $f$ is the focal length, calculated as $\frac{1}{\tan (\theta / 2)}$, while $\theta$ is the UAV FoV value. $z_{\text {n}}$ and $z_{\text {f}}$ are the distances from the UAV to the near and far clipping planes.
 

\textbf{Box object detection algorithm}:
A YOLOv5-based object detection neural network (YbNet) is implemented to extract the moving box location and size from UAV images. The YbNet object detection $\mathcal{Y}(\cdot)$ is defined as
\begin{equation}
\overrightharp{B}_t=\mathcal{Y}(\mathbf{I}_{\text{t}}, \tau_{\text{y}})
\end{equation}
where $\tau_{\text{y}}$ denotes the parameters in the YbNet, The object function of YbNet is to
minimize the euclidean distance between the prediction and groundtruth, which is denoted as
\begin{equation}
\mathcal{L} = \arg \min_{(\tau_{\text{y}})} \sqrt{|\overrightharp{B}_t-\overrightharp{B}_{\text{l}}|^{2}}
\end{equation}
where $\overrightharp{B}_{\text{l}}$ represents the groundtruth of moving box location and size information. By employing the object detection and key points extraction algorithms, the defined semantic information $\mathbf{S}_t$ is obtained as in Eq. (\ref{semantic_ext}).




\subsection{Base Knowledge Definition}
Compared to ImageCom framework that rely on images for 3D scenery construction, our proposed GSCM framework integrates base knowledge, denoted as $\mathcal{B}$, into the semantic communication for metaverse construction. 
In detail, some components in the metaverse, such as the stationary background $O$, remain stable when movable objects $\mathcal{M}$ are in operation. 
These stationary components can be used as base knowledge, transmitted only at the beginning of the metaverse application and thus alleviate the bandwidth problem. 
As depicted in Fig. \ref{wirelesscommunication} (b), the base knowledge is defined to include three parts: 

1) The base scenery information includes the canny image set $\mathcal{V}_c$, extracted from the image set $\mathcal{V}_0$ of the scenery at time $t=0$. These canny images contain information about the metaverse's stationary background and can serve as angle information in the AISG algorithm for 
image generation. 2) The base camera information encompasses the UAV camera parameters $\mathcal{C}$, which include the UAV camera's ID, camera angle, distance, focal length, etc. As shown in Eq. (\ref{construct}), these parameters are used as input for metaverse construction in the Nerf algorithm. 3) The base object information consists of all objects' three-dimensional location coordinates at $t=0$. These coordinates are used to attach objects or scenery within Unity3D, allowing them to be correctly placed.

\begin{algorithm}[t]
	\renewcommand{\algorithmicrequire}{\textbf{Input:}}
	\renewcommand{\algorithmicensure}{\textbf{Output:}}
	\caption{Nerf-based metaverse construction}
        \label{reconstruction}
	\begin{algorithmic}[1]
		\STATE Initialization: Base knowledge $\mathcal{B}$, received moving box data $\overrightharp{B}^{\prime}_t$, received key points data $\overrightharp{K}^{\prime}_t$ 
            \IF{M-AISG}
                \FOR{$i$ $\in$ $\{1, 2\}$}
                \STATE Generate image $\hat{\mathbf{I}}^{\prime}_t$ by Eq. (\ref{sd1}) with input $\mathcal{B}$ and $\overrightharp{K}^{\prime}_t$
                \STATE Generate 3D object $O_i$ by Eq. (\ref{nerf_algorithm}) with input $\hat{\mathbf{I}}^{\prime}_t$ 
                \ENDFOR
                \FOR{each object $O_i$}
                \STATE Attach object $O_i$ in Unity3D.
                \ENDFOR
                
            \ELSE
                \STATE Generate images $\hat{\mathbf{I}}^{\prime}_t$ by Eq. (\ref{sd2}) with $\overrightharp{B}^{\prime}_t$, $\overrightharp{K}^{\prime}_t$, and $\mathcal{B}$.
                \STATE Generate 3D scenery by Eq. (\ref{nerf_algorithm}) with input $\hat{\mathbf{I}}^{\prime}_t$  
                \STATE Attach 3D scenery in Unity3D.

            \ENDIF

		\ENSURE  Metaverse scenery in point cloud formate $\mathbf{P}$
	\end{algorithmic}  
\end{algorithm}

\subsection{NeRF-based Metaverse Construction}
As shown in Fig. \ref{image_generation}, the NeRF-based metaverse construction incorporates both the AISG and a NeRF rendering algorithm. With prompt text descriptions and semantic information, the metaverse is constructed to reflect the physical scenery status in various styles and colors. The whole NeRF-based metaverse construction is detailed in Algorithm \ref{reconstruction}.
\subsubsection{Angle-initialized SD Image Generation Algorithm}
Compared with the general SD model relies solely on text prompt control, the proposed AISG algorithm incorporates a fine-tuned Controlnet that adapts to different spatial control inputs. 
Previous research demonstrates that as the number of input conditions increases, the diffusion model becomes more unstable in generating coherent images \cite{hu2023videocontrolnet}.
Thus, we utilize two distinct approaches to construct the AISG algorithm for image generation. The first approach, named model-based AISG (M-AISG), generates metaverse scenery images in two steps: 1) Using key points as input to generate images of the robotic arm. 2)  Using moving box information and canny images to generate images of the metaverse stationary scenery. The M-AISG generation process is described as
\begin{equation}
\label{sd1}
\hat{\mathbf{I}}_t^{\prime}= \begin{cases}\mathcal{D}\left(W_i, \mathcal{V}_c, \vec{B}_t^{\prime}, \phi_{d_1}\right), & \text { if } i=1 \\ \mathcal{D}\left(W_i, \vec{K}_t^{\prime}, \phi_{d_1}\right), & \text { if } i=2\end{cases},
\end{equation}
where \(W_i\) represents the text prompt defines either to generate images of the robotic arm or the stationary scenery.
The second approach, named as scenery-based AISG (S-AISG), generates multi-angle images of the metaverse scenery with key points, moving boxes location, and canny images together.  The S-AISG generation process is described as
\begin{equation}
\label{sd2}
\hat{\mathbf{I}}_t^{\prime}=\mathcal{D}\left(W, \mathcal{V}_c, \vec{K}_t^{\prime}, \vec{B}_t^{\prime}, \omega_c, \omega_k, \omega_b, \phi_{d_2}\right),
\end{equation}
where \(\omega_{c}\), \(\omega_{k}\), and \(\omega_{b}\) represent the control weights for the canny image, robotic arm, and moving box. These weights determine the clearance of each component in the generated image $\hat{\mathbf{I}}_t^{\prime}$. 
To better illustrate the performance differences between the two AISG approaches, we have named the frameworks model-based GSCM (M-GSCM) and scenery-based GSCM (S-GSCM), which incorporate M-AISG and S-AISG, respectively.


\subsubsection{Nerf Rendering}
The NeRF algorithm renders each image \( \hat{\mathbf{I}}^{\prime}_t \) based on the color value at each pixel. The rendered images \( \bar{\mathbf{I}}^{\prime}_t \) are obtained to
aggregate radiance along a ray emitted from the $i$-th UAV position through pixel $p$. This rendering process is denoted as
\begin{equation}
\label{nerf_algorithm}
\bar{\mathbf{I}}_t^{\prime}=\int_{l_s}^{l_e} T(l) \sigma(\mathbf{r}(l)) \mathbf{c}(\mathbf{r}(l), \vec{d}) d l
\end{equation}
where the function \( T(l) = \exp\left(-\int_{l_s}^{l} \sigma(\mathbf{r}(\mu)) \, d\mu\right) \) denotes the accumulated transmittance along the ray, quantifying the likelihood that a ray travels from position \( l_s \) to \( l \) without encountering any particles. The function \( \mathbf{c}(\mathbf{r}(l), \overrightharp{d}) \) calculates the color at a specific position $l$ along the ray, influenced by the direction vector \( \overrightharp{d} \) of the ray, and the variable \( \sigma(\mathbf{r}(l)) \) represents the density of the scene at position $l$ along the ray. 

\begin{figure*}[t]
\centering
\subfigure[Key point error.]
{
\includegraphics[width=5.43cm]{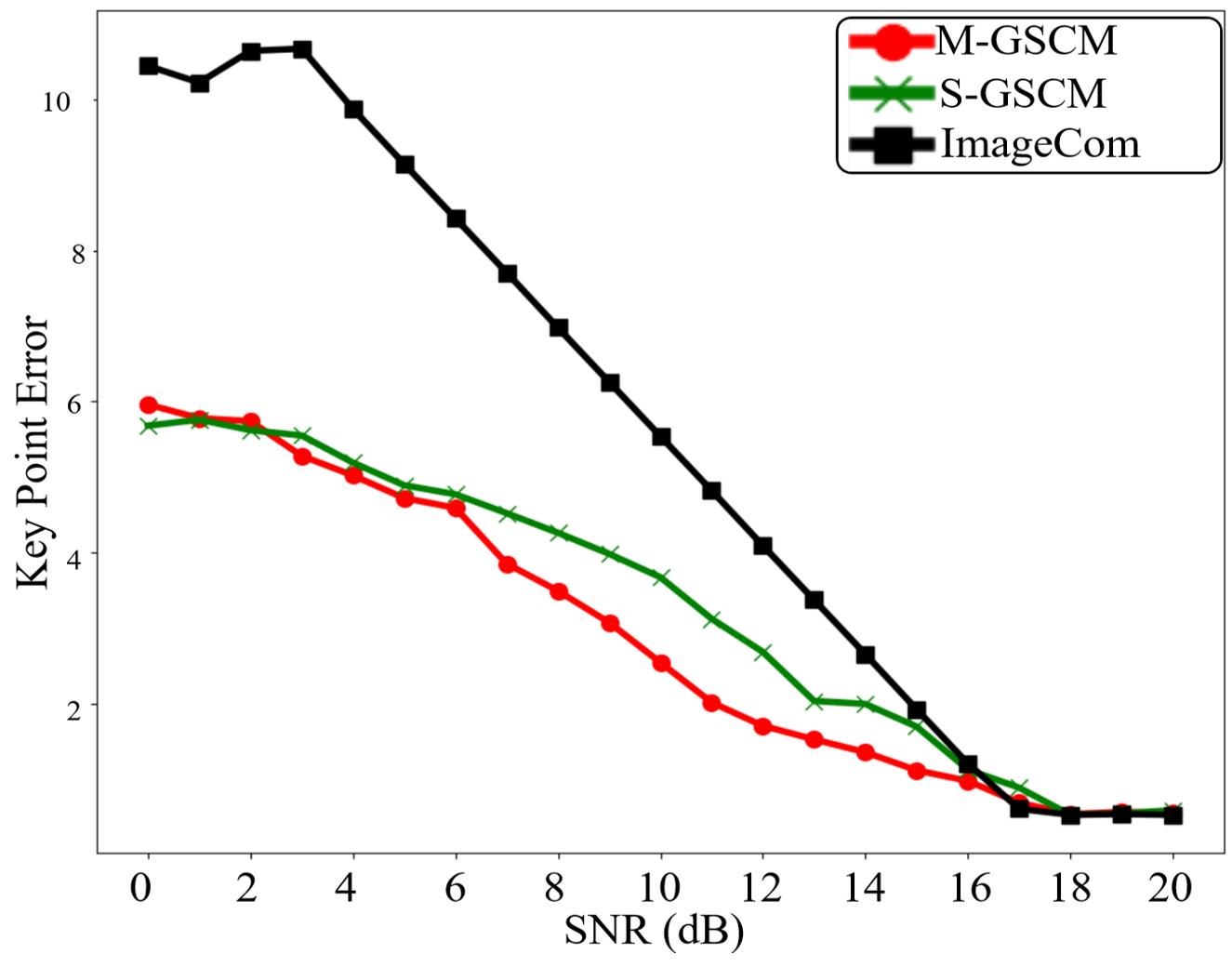}
}
\subfigure[Point-to-point.]{
\includegraphics[width=5.5cm]{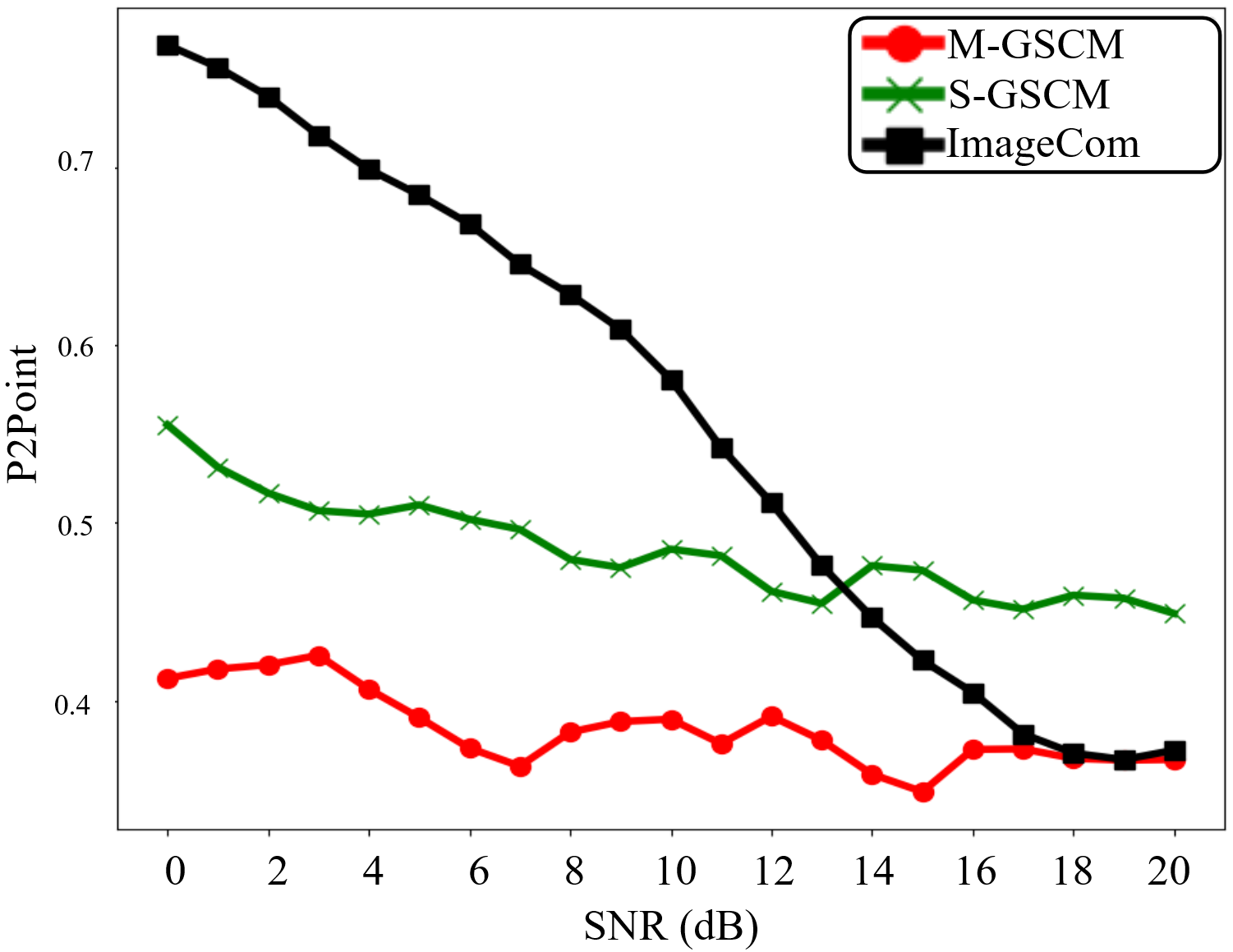}
}
\subfigure[Transmission delay.]{
\includegraphics[width=5.5cm]{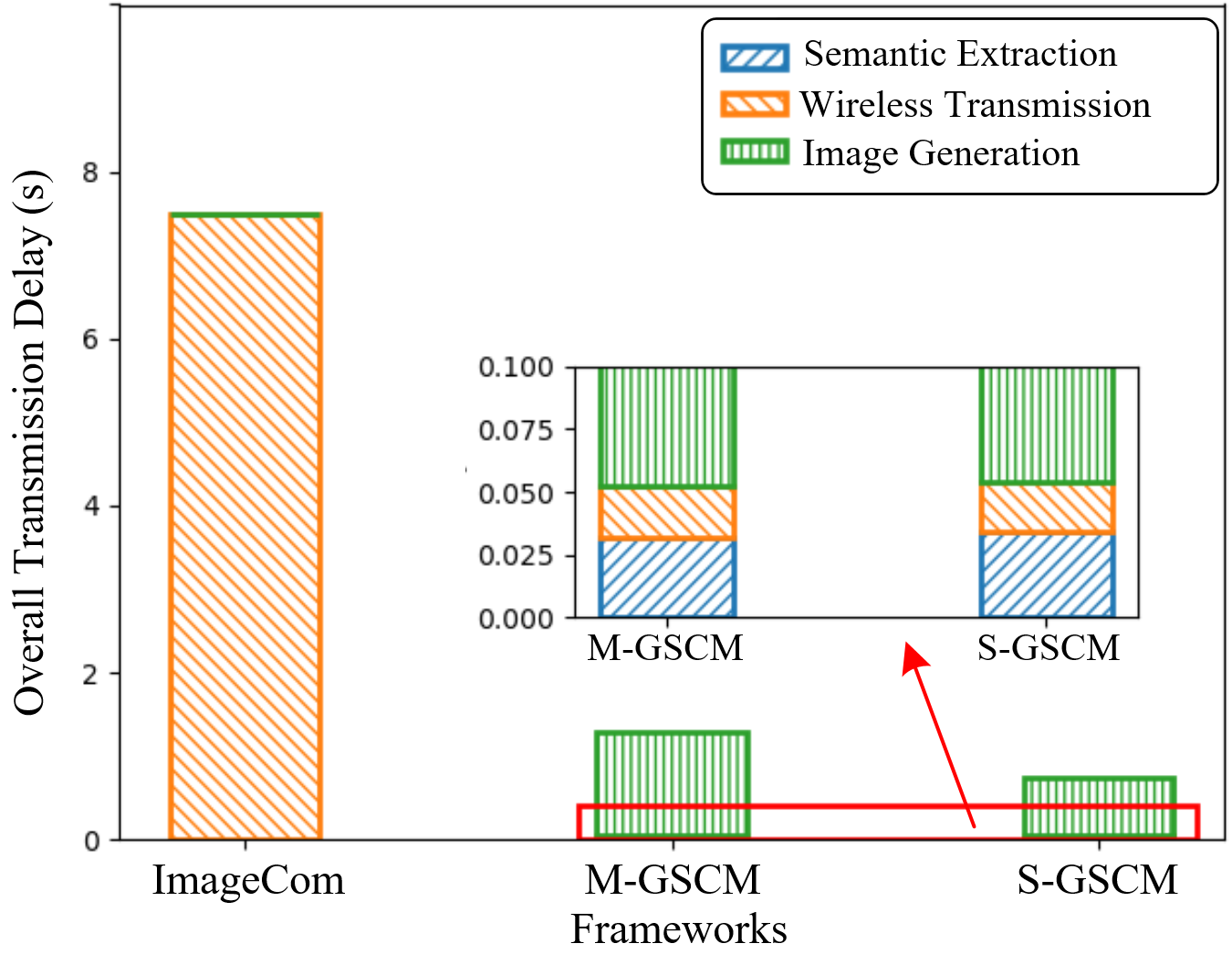}
}

\caption{Experiment performance of key point error, point-to-point, and time delay}
\label{ex22}
\end{figure*}

\section{Experiments}
In this section, we present the experimental platform, evaluation metrics, and a comparative analysis of proposed GSCM against ImageCom framework to verify effectiveness.

\subsection{Experiment Platform}
The experimental platform utilizes Python 3.11.0, PyTorch 2.1, and Unity3D on Ubuntu 18.04, supported by four RTX 4090 GPUs.



\subsection{Evaluation Metrics}
We evaluate goal-level performance using three different metrics to compare 1-D to 3-D information between the transmitter and edge, including key point error (KPE), point-to-point (P2Point), and transmission delay (TD).



\textbf{KPE}: The goal of metaverse construction is to accurately represent the operational status of all virtual counterparts according to physical objects. Therefore, KPE is used to calculate the pose and position differences of the robotic arm and moving box between the transmitter and the edge.


\textbf{P2Point}: 
Both the transmitter and the edge's metaverse scenery are operated in Unity3D platform that enables point cloud output using FM Points plugin, as depicted in \cite{wang2023task}. 
In a clear metaverse environment, point clouds tend to cluster rather than disperse to clearly illustrate objects. Therefore, P2Point is employed to assess the viewing experience of clients in the metaverse by calculating the point cloud geometric differences between the transmitter and the edge.

\textbf{TD}: \textcolor{black}{The transmission delay is used to estimate client viewing experience, which comprises semantic information extraction, wireless transmission, and image generation.}

\subsection{Experiment Platform}

Fig. \ref{ex22} (a) plots the KPE between the transmitter and edge under wireless channel with different SNR conditions. The error is calculated using the pixel distance, assuming $d = 10$ pixels. A lower KPE indicates a higher accuracy of the metaverse content's pose generated at the edge side, and the overall KPE result is ranked as M-GSCM$\textless$S-GSCM$\textless$ImageCom. 
Specifically, 
all the frameworks shown an decrease with the SNR increase, indicating the metaverse content pose is affected by the wireless channel conditions.
Compared to the ImageCom, both the M-GSCM and S-GSCM frameworks achieves a lower KPE value for SNR below 17 dB, and reduces the KPE by 42.3\%  and 45.6\% at 0 dB. 
This indicates that with illogical conditions input, the pretrained AISG model in the GSCM framework will generate metaverse robots appearing in incorrect poses. However, the robots remain easily distinguishable. Meanwhile, the ImageCom framework generates excessive blurring in images, which prevents the identification of metaverse robots posture in the images after wireless communication.
Fig. \ref{ex22} (b) plots the P2Point results after smoothing, revealing the geometric differences in 3D scenery between the edge and transmitter. A lower P2Point value indicates clearer metaverse scenery construction after wireless communication. It can be seen that the overall P2Point is ranked as M-GSCM$\textless$S-GSCM$\textless$ImageCom. 
Specifically, the M-GSCM and S-GSCM achieve an 44.7\% and 29.5\% improvement compared to ImageCom frameworks, respectively, at 0 dB conditions. 
This is because, although the proposed GSCM framework might inaccurately position the robotic arm, it can still generate accurate and stable surrounding sceneries with base information, thereby maintaining the overall 3D scenery with less random points. In contrast, the ImageCom framework generate distorted 3D scenery, significantly disturbs the user's viewing experience.

Fig. \ref{ex22} (c) plots the semantic information extraction, wireless transmission, and image generation delay of our proposed GSCM frameworks compared to ImageCom framework. 
It can be seen that the transmission delay is ranked as S-GSCM$\textless$M-GSCM$\textless$ImageCom.
The M-GSCM and S-GSCM frameworks achieve significant improvement by transmitting only key points and boxes information, compared with the extensive data required for image pixels in ImageCom framework. Although the GSCM incorporate additional steps for semantic information extraction and image generation, each frame only requires around 0.67 seconds to generate images at the edge. 
This indicates a significant improvement in wireless communication efficiency for the M-GSCM and S-GSCM compared with the ImageCom, at about 81.4\% and 92.6\% respectively.

\section{Conclusion}
This paper proposes a goal-oriented semantic communication framework for metaverse application (GSCM) to enhance the effectiveness and efficiency of wireless metaverse construction. By incorporating semantic information, innovative angle-initialized stable diffusion image generation algorithm, and the NeRF algorithm into metaverse construction, our proposed GSCM reduces transmission delay by 92.6\% and improves the accuracy of virtual object operation status and metaverse construction clearance by 45.6\% and 44.7\%, respectively, compared to traditional image communication framework.
\small
\bibliographystyle{IEEEtran}
\bibliography{egbib}
\end{document}